\documentclass[11pt]{article}

\usepackage{amsmath,amssymb,amsthm}
\usepackage[margin=2.5cm]{geometry}
\usepackage{hyperref}
\usepackage{color}

\newcounter{Theorems}
\newcounter{Definitions}
\newcounter{Conjectures}
\newcommand{\p}{\partial}

\newcommand{\Oc}{{\mathcal O}}

\renewcommand{\a}{{\alpha}}

\newcommand{\g}{{\gamma}}

\renewcommand{\l}{{\lambda}}

\renewcommand{\t}{{\theta}}

\newcommand{\omegaLLL}{{\Omega_L{}^L_L}}
\newcommand{\omegaLRR}{{\Omega_L{}^R_R}}
\newcommand{\omegaRLL}{{\Omega_R{}^L_L}}
\newcommand{\omegaRRR}{{\Omega_R{}^R_R}}
\newcommand{\xiLLL}{{\Xi_L{}^L_L}}
\newcommand{\xiLRR}{{\Xi_L{}^R_R}}
\newcommand{\xiRLL}{{\Xi_R{}^L_L}}
\newcommand{\xiRRR}{{\Xi_R{}^R_R}}

\begin{document}
\begin{titlepage}
\begin{flushright}

\end{flushright}

\begin{center}
{\Large\bf $ $ \\ $ $ \\
Normal form of nilpotent vector field near the tip of the pure spinor cone
}\\
\bigskip\bigskip\bigskip
{\large Andrei Mikhailov and Dennis Zavaleta}
\\
\bigskip\bigskip
{\it Instituto de Fisica Teorica, Universidade Estadual Paulista\\
R. Dr. Bento Teobaldo Ferraz 271, 
Bloco II -- Barra Funda\\
CEP:01140-070 -- Sao Paulo, Brasil\\
}

\vskip 1cm
\end{center}

\begin{abstract}
Pure spinor formalism implies that supergravity equations in space-time are equivalent
                 to the requirement that the worldsheet sigma-model satisfies certain properties.
                 Here we point out that one of these
                 properties has a particularly transparent geometrical interpretation. 
                 Namely, there exists an odd nilpotent vector field on some singular supermanifold,
                 naturally associated to space-time. All supergravity fields are
                 encoded in this vector field, as coefficients in its normal form. The nilpotence
                 implies, modulo some zero modes, that they satisfy the SUGRA equations of motion.
\end{abstract}

\vfill
{\renewcommand{\arraystretch}{0.8}%
}

\end{titlepage}

\tableofcontents

\section{Introduction}\label{Introduction}

In the low energy limit of superstring theory,
spacetime fields satisfy supergravity (SUGRA) equations of motion,
which are super-analogues of the Einstein equations.
It is one of the main principles of string theory, that these target space equations of motion
are equivalent to the BRST invariance of the string worldsheet theory. When they are satisfied,
the space of fields is an infinite-dimensional $Q$-manifold
(a manifold with an odd nilpotent vector field \cite{Alexandrov:1995kv}).
But in the case of \emph{pure spinor string}, the sigma-model also defines a \emph{finite-dimensional} $Q$-manifold.
Indeed,
the action of the BRST operator on matter fields and pure spinor ghosts does not contain worldsheet derivatives.
(The worldsheet derivatives will appear when we consider the action on the conjugate momenta
     to matter fields and pure spinor ghosts, but they can be considered separately.)
This means that, if we think of the pure spinor ghosts as part of target space, the BRST operator defines
\emph{on the target space} an odd nilpotent vector field, which we denote $Q$. In other words,
the target space of the pure spinor sigma-model (a finite-dimenisional supermanifold) is a $Q$-manifold.
Moreover, in generic space-time (for example in $AdS_5\times S^5$, but not in flat space-time)
the energy-momentum tensor and the $b$-ghost can also be interpreted as symmetric tensors
on the target space (see \cite{Mikhailov:2017mdo}).

How to classify a generic odd nilpotent vector field $Q$? A vector field can usually be ``simplified'' by
a clever choice of coordinates. This is called  ``normal form''.
If a vector field is non-vanishing,
one can choose coordinates so that the it is $\partial\over\partial\theta$ where $\theta$ is
one of fermionic coordinates. If $Q$ vanishes at some point, then the normal form would be
(in the notations of \cite{Alexandrov:1995kv}) $\eta^a {\partial\over\partial x^a}$.
But in out case, the target space is not a smooth supermanifold, because pure spinor ghosts live on a cone.
The vector $Q$ vanishes precisely at the singular locus, and the problem of classification of normal forms is
a nontrivial cohomological computation. This is what we will do in this paper. We will find that
the space of equivalence classes of odd nilpotent vector fields in a vicinity of the singular locus
is equivalent to the space of the classical SUGRA solutions.
This is true modulo some \emph{``zero modes''} --- a finite-dimensional subspaces of soultions (see \cite{Mikhailov:2014qka})
which we ignore in this paper.

Some details of our computations can be found in the
\href{https://andreimikhailov.com/math/vector-fields/index.html}{\textbf{\textcolor{blue}{HTML version of this paper}}}.

\subsection{Definition of $M$}\label{sec:IntroM}

The particular singularity which we are interested in can be described as follows.
Consider the space $M$ with bosonic coordinates $x^m$ ($m$ running from 1 to 10) and
$\lambda_L^{\alpha}$, $\lambda_R^{\hat{\alpha}}$ ($\alpha$ and $\hat{\alpha}$ both running from 1 to 16),
and fermionic $\theta_L^{\alpha}$ and $\theta_R^{\hat{\alpha}}$, with the constraint:
\begin{equation}
     \lambda_L^{\alpha}\Gamma_{\alpha\beta}^m\lambda_L^{\beta} =
     \lambda_R^{\hat{\alpha}}\Gamma_{\hat{\alpha}\hat{\beta}}^m\lambda_R^{\hat{\beta}} = 0
     \label{IntroPSCone}\end{equation}
where $\Gamma^m$ are ten-dimensional gamma-matrices. These constraints are called ``pure spinor constraints''.
We understand Eqs. (\ref{IntroPSCone}) as specifying the singular
locus in $M$, from the point of view  of differential geometry. All we need from these
equations is to know how $M$ deviates from being smooth.
The singular locus is the tip of the cone (\ref{IntroPSCone}):
\begin{equation}
   \lambda_L = 0 \mbox{ \tt\small or } \lambda_R = 0
   \label{SingularityLocus}\end{equation}
Pure spinor constraints (\ref{IntroPSCone}) are invariant under the action of the group
\begin{equation}
     G = {\rm Spin}(10)_L\times {\bf C}^{\times}_L \times {\rm Spin}(10)_R\times {\bf C}^{\times}_R
     \label{DefG}\end{equation}
The diagonal
\begin{equation}
     {\bf C}^{\times}\subset {\bf C}_L^{\times}\times {\bf C}_R^{\times}
     \label{GhostNumber}\end{equation}
is called ``ghost number symmetry''. Infinitesimal ghost number symmetry is generated by
$\lambda_L^{\alpha}{\partial\over\partial\lambda_L^{\alpha}} +
                   \lambda_R^{\hat{\alpha}}{\partial\over\partial\lambda_R^{\hat{\alpha}}}$.
                          
Consider an odd vector field $Q$ satisfying the following properties:
\begin{itemize}\item $Q$ has ghost number 1, \textit{i.e.}:
        \begin{equation}
           \left[\lambda_L^{\alpha}{\partial\over\partial\lambda_L^{\alpha}} +
                           \lambda_R^{\hat{\alpha}}{\partial\over\partial\lambda_R^{\hat{\alpha}}}\;,\;Q\right] = Q
           \label{QhasGhostNumberOne}\end{equation}
          \item $Q^2 = 0$\item $Q$ is ``smooth'' in the sense that it can be
          obtained as a restriction to the cone (\ref{IntroPSCone}) of a smooth (but not nilpotent)
          vector field
          in the space parametrized by unconstrained $x,\theta,\lambda$\item $Q$ is zero at $\lambda_L = \lambda_R = 0$ \end{itemize}
We want to classify such vector fields modulo coordinate transformations.
Coordinate transformations are supermaps $(x,\lambda,\theta)\mapsto (\tilde{x},\tilde{\lambda},\tilde{\theta})$
such that $\tilde{\lambda}$ satisfy the same constraints (\ref{IntroPSCone}).

Such a vector field is one of the geometrical structures associated to the pure spinor
superstring worldsheet theory \cite{Berkovits:2001ue},\cite{Guttenberg:2008ic}.
In particular, flat background (empty ten-dimensional spacetime) corresponds to $Q=Q^{\rm flat}$:
\begin{align} Q^{\rm flat} \;=\;
 &Q^{\rm flat}_L + Q^{\rm flat}_R \mbox{ \tt\small where: }\label{QFlat} \\ Q_L^{\rm flat}\;=\;
 &\lambda^{\alpha}_L{\partial\over\partial\theta^{\alpha}_L} +
                   (\lambda^{\alpha}_L\Gamma_{\alpha\beta}^m\theta^{\beta}_L)
                   {\partial\over\partial x^m}\nonumber{} \\ Q_R^{\rm flat}\;=\;
 &\lambda^{\hat{\alpha}}_R{\partial\over\partial\theta^{\hat{\alpha}}_R} +
                    (\lambda^{\hat{\alpha}}_R\Gamma_{\hat{\alpha}\hat{\beta}}^m\theta^{\hat{\beta}}_R)
                    {\partial\over\partial x^m}\nonumber{} \end{align}
String worldsheet
theory also has, besides $Q$, some other structures which are less geometrically transparent
(various couplings in the string worldsheet sigma-model). All these structures should satisfy certain
consistency conditions. 
\begin{itemize}\item 
        \emph{Question:} is it true, that just a nilpotent vector field $Q$ already
        includes, as various coefficients in its normal form, all the supergravity fields, and the supergravity
        equations of motion are automatically satisfied (\textit{i.e.} follow from $Q^2 = 0$)?
        \end{itemize}
This may be false in two ways. First, it could be that some supergravity fields do not
enter as coefficients in the normal form of $Q$
(\textit{i.e.} they would only appear as some couplings in the sigma-model, but would not enter in $Q$).
Second, it could be that just $Q^2 = 0$ would not be enough to impose SUGRA equations
of motion (\textit{i.e.} one would have to also require the $Q$-invariance of the
                  worldsheet sigma-model action).

\subsection{Our results}\label{sec:IntroResults}

In this paper we will derive the normal form of $Q$ as a deformation of $Q^{\rm flat}$:
\begin{equation}
     Q = Q^{\rm flat} + \epsilon Q_1 + \epsilon^2 Q_2 + \ldots
     \label{ExpansionOfQAroundFlatSpace}\end{equation}
Our analysis will be restricted to the terms linear in $\epsilon$ (\textit{i.e.} $Q_1$).
It turns out that $Q_1$ is parameterized by some tensor fields satisfying certain
hyperbolic partial differential equations. These fields are in one-to-one correspondence
with the fields of the Type II SUGRA, and our hyperbolic equations are the equations of motion
of the linearized Type II SUGRA.

It is useful to compare to the pure spinor description of the super-Yang-Mills equations.
The super-Yang-Mills equations
are equivalent to having an odd nilpotent operator:
\begin{equation}
     Q_{\rm SYM} =
     \lambda^{\alpha}
     \left(
           {\partial\over\partial\theta^{\alpha}} +
           \Gamma_{\alpha\beta}^m \theta^{\beta}{\partial\over\partial x^m}
           + A^a_{\alpha}(x,\theta){\bf t}_a
           \right)
     \label{QSYM}\end{equation}
where ${\bf t}_a$ are generators of the gauge group, and $A^a_{\alpha}(x,\theta)$ is vector potential.
Zero solution corresponds to $A_{\alpha} = 0$. In this sense, the SYM solutions can be
considered as deformations of the differential operator:
\begin{equation}
   Q_{\rm SYM}^{(0)} = 
     \lambda^{\alpha}
     \left(
           {\partial\over\partial\theta^{\alpha}} +
           \Gamma_{\alpha\beta}^m \theta^{\beta}{\partial\over\partial x^m}
           \right)
     \end{equation}
where the leading symbol (\textit{i.e.} the derivatives) remains undeformed.
Here we consider, instead, the deformations of the leading symbol.

\subsection{Relation to partial $G$-structures}\label{GStructures}

The variables $\lambda_L$ and $\lambda_R$ parametrize the normal direction to the
singularity locus $Z\subset M$:
\begin{equation}
   i\;:\; Z\rightarrow M
   \end{equation}
The first infinitesimal neighborhood is a bundle
over $Z$ with the fiber $C_L\times C_R$ --- the product of two cones.
Filling the cones, we obtain a vector bundle over $Z$ with the fiber $V = {\bf C}^{32}$.
The vector field $Q$ is power series in $\lambda_L,\lambda_R$, with zero at the tip of $C_L\times C_R$.
The derivative of $Q$ at the zero locus defines a linear map:
\begin{equation}
   Q_*\;:\; V\rightarrow i^* TM
   \end{equation}
This map is not an isomorphism, since the image of $Q_*$ only covers a $(0|32)$-dimensional
subbundle of $TZ$. We can interpret $M$ as  $(C_L\times C_R)\times_G \widehat{Z}$
where $\widehat{Z}$ is a partial frame bundle of $Z$ and  $G$ is given by Eq. (\ref{DefG}).
It was shown in \cite{Mikhailov:2015sva} that $Q$ defines a connection in a partial
$G$-structure on $Z$ with some constraints on torsion, modulo some equivalence relation.

The relation to previous work on SUGRA constraints 
\cite{Nilsson:1981bn},
\cite{Howe:1983sra},
\cite{Witten:1985nt},
\cite{Shapiro:1986xp}, 
\cite{Chau:1988sm}, 
\cite{Bergshoeff:1990mr},
\cite{Bergshoeff:1991ei}, 
\cite{Howe:1991bx},
\cite{Howe:1991mf}
can be established along these lines.

\subsection{Divergence of a nilpotent vector field}\label{sec:Divergence}

Let us fix some volume form $\mbox{vol}$, an integral form on $M$.
Then we can consider, for any vector field $V$, its divergence $\mbox{div}\, V$.
This is by definition the Lie derivative of $\mbox{vol}$ along $V$:
\begin{equation}
   {\cal L}_V\mbox{vol} = (\mbox{div}\,V)\mbox{vol}
   \end{equation}
In particular, for a nilpotent odd vector field $Q$, we can consider the cohomology class:
\begin{equation}
   [\mbox{div}\,Q] = \mbox{div}\,Q \quad \mbox{mod}\quad Q(\ldots)
   \end{equation}
This cohomology class does not depend on the choice of the volume form $\mbox{vol}$.

The cohomology class $[\mbox{div}\;Q]$ plays two importan roles in our approach.
First, they allow to reduce the study of $Q$ to the first infinitesimal neighborhood of the singularity
locus given by Eq. (\ref{SingularityLocus}). Second, it allows to prove that there is no obstacle
to extending the linearized deformations (the term $\epsilon Q_1$ in Eq. (\ref{ExpansionOfQAroundFlatSpace}))
to higher orders in $\epsilon$. We will now explain this.

\subsubsection{Requiring $\mbox{div}\,Q = 0$}\label{sec:FirstInfinitesimalNeighborhood}

We required that $Q$ has ghost number one, see Eq. (\ref{QhasGhostNumberOne}). But the ``ghost number''
$\lambda_L^{\alpha}{\partial\over\partial\lambda_L^{\alpha}} +
             \lambda_R^{\hat{\alpha}}{\partial\over\partial\lambda_R^{\hat{\alpha}}}$
is coordinate-dependent. It is not invariant under a change of coordinates:
\begin{equation}
   \lambda^{\alpha} \mapsto \mbox{polynomial of }\; \lambda
   \end{equation}
In fact, we can relax Eq. (\ref{QhasGhostNumberOne}) by replacing it with the following two requirements:
\begin{align}  
 &Q = 0 \quad \mbox{when} \quad\lambda_L = 0\; \mbox{ or }\; \lambda_R = 0\label{QisZeroAtLocus} \\  
 &[\mbox{div}\;Q] = 0\label{ZeroDiv} \end{align}
Indeed Eq. (\ref{QisZeroAtLocus}) implies that $Q$ is a vector field of ghost number one plus
vector fields of ghost numbers two and higher. 
Although we have not computed $H^{>1}\left(\left[Q^{\rm flat},\_\right]\right)$,
the results of \cite{Mikhailov:2014qka} suggest that
\begin{equation}
   \mbox{div} \;:\;
   H^2\left(\left[Q^{\rm flat},\_\right]\right) \longrightarrow
   H^2\left(Q^{\rm flat},\mbox{functions}\right)
   \end{equation}
is an isomorphism, and $H^{>2}\left(\left[Q^{\rm flat},\_\right]\right)$ is zero modulo finite-dimensional spaces.
Then, Eq. (\ref{ZeroDiv}) implies that the terms of the ghost number higher than one in $Q$ can be removed
by the coordinate redefinition. In other words, the normal form of $Q$ can be always choosen to be of the ghost nuber one.
It is enough to study the first infinitesimal neighborhood of the singularity locus.

\subsubsection{Vanishing of obstacles to nonlinear solution}\label{sec:VanishingOfObstacles}

It is necessary to extend this analysis to full nonlinear SUGRA equations,
\textit{i.e.} higher order terms in Eq. (\ref{ExpansionOfQAroundFlatSpace}).
The potential obstacle to extending linearized solutions to the solution of the
nonlinear equation $Q^2=0$ lies in $H^2\left(\left[Q^{\rm flat},\_\right]\right)$.
We will not compute $H^2\left(\left[Q^{\rm flat},\_\right]\right)$ in this paper,
but the results of \cite{Mikhailov:2014qka} suggest that
$H^2\left(\left[Q^{\rm flat},\_\right]\right)$
is actually nonzero.
(We would expect it to be roughly isomorphic to $H^1\left(\left[Q^{\rm flat},\_\right]\right)$ which we compute here.)
But we also know that the \emph{actual} obstacle is zero, because of the consistency of the nonlinear supergravity
equations of \cite{Berkovits:2001ue}. This means that $\{Q_1,Q_1\}$ must be a coboundary.
In our language, this can be proven in the following way. 
Let us choose $\mbox{vol}$ so that the divergence of $Q^{\rm flat}$ is zero.
The divergence of $Q_1$, and therefore of $\{Q_1,Q_1\}$,
is $Q_0$-exact (this statement does not depend on the choice of $\mbox{vol}$). This is because $\mbox{div} \,Q_1$ has
ghost number 1. The cohomology at ghost number 1 is finite-dimensional, and in fact those $Q_1$ with nonzero
$\mbox{div}\,Q_1 \; \mbox{mod}\; Q_0(\ldots)$ are non-physical
(see \cite{Mikhailov:2014qka} and references there).
At the same time, the divergence of the elements of $H^2\left(\left[Q^{\rm flat},\_\right]\right)$
is nonzero.  Therefore the obstacle actually vanishes.

\section{Notations}\label{IntroNotations}

To avoid the discussion of reality conditions, we consider complex vector fields. The notation:
\begin{equation}
   {\bf C}\langle v_1,v_2,\ldots \rangle
   \label{LinearSpan}\end{equation}
means the space of all linear combinations of vectors $v_1,v_2,\ldots$ with complex coefficients.

We introduce the abbreviated notations:
\begin{align} (\!(\lambda\theta)\!)^m \;=\;
 &\lambda^{\alpha}\Gamma^m_{\alpha\beta}\theta^{\beta}\nonumber{} \\ (\!(\lambda\theta\theta)\!)_{\gamma} \;=\;
 &\lambda^{\alpha}\Gamma^m_{\alpha\beta}\theta^{\beta}\theta^{\delta}\Gamma^{m}_{\delta\gamma}\nonumber{} \\ [v\otimes\psi]^{1/2}_{\alpha} \;=\;
 &\Gamma^m_{\alpha\beta} v_m \psi^{\beta}\nonumber{} \\ [v\otimes\psi]_{1/2}^{\alpha} \;=\;
 &\Gamma^{m\alpha\beta} v_m \psi_{\beta}\nonumber{} \end{align}

\section{Setup for cohomological perturbation theory}\label{CohomologicalPerturbationTheory}

\subsection{Definition of $\theta^{\alpha}_L$ and $\theta^{\hat{\alpha}}_R$}\label{sec:DefThetas}

We \emph{define odd coordinates} $\theta$ so that:
\begin{equation}
     Q_L\theta_L^{\alpha} = \lambda_L^{\alpha} + O(\theta^2),\quad
     Q_L\theta_R^{\hat{\alpha}} = O(\theta^2),\quad
     Q_R\theta_R^{\hat{\alpha}} = \lambda_R^{\hat{\alpha}} + O(\theta^2),\quad
     Q_R\theta_L^{\alpha} = O(\theta^2)
     \label{DefTheta}\end{equation}

\subsection{Flat $Q$ and expansion around it}\label{sec:FlatQ}

Flat spacetime corresonds to $Q=Q^{\rm flat} = Q^{\rm flat}_L + Q^{\rm flat}_R$ where:
\begin{align} Q_L^{\rm flat}\;=\;
 &\lambda_L{\partial\over\partial\theta_L} + (\!(\lambda_L\theta_L)\!){\partial\over\partial x}\nonumber{} \\ Q_R^{\rm flat}\;=\;
 &\lambda_R{\partial\over\partial\theta_R} + (\!(\lambda_R\theta_R)\!){\partial\over\partial x}\nonumber{} \end{align}
Let us consider $Q$ as a small deformation of $Q^{\rm flat}$:
\begin{equation}
   Q = Q^{\rm flat} + \epsilon Q_1
   \label{LinearizedQ}\end{equation}
to the first order in $\epsilon$. Such deformations form a linear space. They correspond to
odd vector fields $Q_1$ satisfying:
\begin{equation}
   [Q^{\rm flat} , Q_1] = 0
   \end{equation}
modulo the equivalence relation, corresponding to the action of diffeomorphisms:
\begin{equation}
   Q_1\simeq Q_1 + [Q^{\rm flat},R]
   \end{equation}
where $R$ is a ghost number zero vector field on $M$.
Therefore, the classification of nilpotent vector fields of the form (\ref{LinearizedQ})
is equivalent to the computation of the
cohomology of the operator $[Q^{\rm flat},\_]$ on the space of vector fields.

In the rest of this paper we will compute the cohomology of  $[Q^{\rm flat},\_]$ on the space
of vector fields.

\subsection{Spectral sequence}\label{sec:SpectralSequence}

The grading operator:
\begin{equation}
   N = \theta_L {\partial\over\partial \theta_L} +\theta_R {\partial\over\partial\theta_R}
   + \lambda_L {\partial\over\partial\lambda_L} + \lambda_R {\partial\over\partial\lambda_R}
   \label{GradingOperator}\end{equation}
defines a filtration on the algebra of functions on $\mbox{Fun}(M)$,
and  on the space of vector fields as a $\mbox{Fun}(M)$-module. Let $F^N\mbox{Vect}$
be the space of vector fields with grade at least $N$. This filtration defines a spectral sequence converging to the cohomology of $[Q^{\rm flat},\_]$.

\subsection{First page}\label{sec:FirstPage}

The first page of this spectral sequence is the cohomology of:
\begin{equation}
   [Q^{(0)}_L + Q^{(0)}_R\,,\,\_] =
   \left[\lambda_L{\partial\over\partial\theta_L} + \lambda_R{\partial\over\partial\theta_R}\,,\,\_\right]
   \end{equation}
on the space of vector fields on $M$. For a set of coordinates $x,y,\ldots$ we denote
$\mbox{Fun}(x,y,\ldots)$ the space of functions of $x,y,\ldots$ and $\mbox{Vect}(x,y,\ldots)$
the space of vector fields (\textit{i.e.} differentiations of $\mbox{Fun}(x,y,\ldots)$).
Let us introduce the following complexes:
\begin{align} C_L^{\rm vect}\;=\;
 &\mbox{$\mbox{Vect}(\theta_L,\lambda_L)$ with differential $[Q_L^{(0)},\_]$}\nonumber{} \\ C_R^{\rm vect}\;=\;
 &\mbox{$\mbox{Vect}(\theta_R,\lambda_R)$ with differential $[Q_R^{(0)},\_]$}\nonumber{} \\ C_L^{\rm fun}\;=\;
 &\mbox{$\mbox{Fun}(\theta_L,\lambda_L)$ with differential $Q_L^{(0)}$}\nonumber{} \\ C_R^{\rm fun}\;=\;
 &\mbox{$\mbox{Fun}(\theta_R,\lambda_R)$ with differential $Q_R^{(0)}$}\nonumber{} \end{align}
Then, $\mbox{Vect}(M)$ with differential $Q_L^{(0)} + Q_R^{(0)}$ decomposes as follows:
\begin{align} \mbox{Vect}(M) \;=\;
 &\phantom{\;\oplus\;} \mbox{Fun}(x) \otimes C_R^{\rm fun}\otimes C_L^{\rm vect}\label{FunXCRCL} \\  
 &\;\oplus\;\mbox{Fun}(x) \otimes C_L^{\rm fun}\otimes C_R^{\rm vect}\label{FunXCLCR} \\  
 &\;\oplus\;\mbox{Fun}(x) \otimes C_L^{\rm fun}\otimes C_R^{\rm fun} \otimes {\partial\over\partial x}\label{FunXCLCRDx} \end{align}
(We do not need to take care about the completions of the tensor products, since all our functions
    are polynomials in $\theta$ and $\lambda$.)
The cohomology of $C_L^{\rm fun}$ and $C_R^{\rm fun}$ is well known,
see \textit{e.g.} the review part of \cite{Mafra:2009wq}:
\begin{align} H^0(C^{\rm fun})\;=\;
 &{\bf C}\langle {\bf 1}\rangle\nonumber{} \\ H^1(C^{\rm fun})\;=\;
 &{\bf C}\left\langle
                  (\!(\lambda\theta)\!),\;[(\!(\lambda\theta)\!)\otimes \theta]^{1/2}
                  \right\rangle\nonumber{} \\ H^2(C^{\rm fun})\;=\;
 &{\bf C}\Big\langle
                  (\!(\lambda\theta)\!)^m(\!(\lambda\theta)\!)^n\Gamma_{mn}\theta,\;
                  (\!(\lambda\theta)\!)^m(\!(\lambda\theta)\!)^n(\theta\Gamma_{lmn}\theta)
                  \Big\rangle\nonumber{} \\ H^3(C^{\rm fun})\;=\;
 &{\bf C}\left\langle
                  (\!(\lambda\theta)\!)^l(\!(\lambda\theta)\!)^m(\!(\lambda\theta)\!)^n (\theta \Gamma_{lmn}\theta)
                  \right\rangle\nonumber{} \end{align}
Parts of the cohomology of $C_L^{\rm vect}$ and $C_R^{\rm vect}$ which are relevant to this work
will be computed in Section \ref{CohomologyQ0}.

\section{Cohomology of $Q^{(0)}$ in the space of vector fields}\label{CohomologyQ0}

\subsection{Notations}\label{sec:Notations}

Let $X$ denote the singular supermanifold parametrized by bosonic $\lambda^{\alpha}$ and fermionic
$\theta^{\alpha}$ satisfying the pure spinor constraint:
\begin{equation}
     \lambda^{\alpha}\Gamma_{\alpha\beta}^m \lambda^{\beta} = 0
     \label{NotationsPureSpinorConstraint}\end{equation}
(The space $M$ introduced in 
     Section \ref{sec:IntroM}
     is the direct product of two copies of $X$, and the space parametrized by $x^m$.)
Let ${\cal O}(X)$ denote the algebra of polynomial functions on $X$, and  $Vect(X)= Der(\Oc (X))$
the space of polynomial vector fields. Consider the odd nilpotent vector field $Q^{(0)}$:
\begin{equation}
   Q^{(0)} = \lambda^{\alpha}{\partial\over\partial\theta^{\alpha}}
   \end{equation}
The commutation $[Q^{(0)},-]$ is a nilpotent operator on  $Vect(X)$.
We will now compute the cohomology of this operator.

Any vector field $V\in  Vec(X)$ can be written as
\begin{align} V \;=\;
 &\xi^{\alpha}(\lambda,\theta) \frac{\p }{\p \l ^\a }  +  u^{\alpha}(\lambda,\theta)\frac{\p }{\p \t ^\a}\nonumber{} \\  
 &(\l \g _m)_\a  \xi^{\alpha}  = 0\nonumber{} \end{align}
The condition $(\l \g _m)_\a  \xi^{\alpha}  = 0$ is needed because $\lambda^{\alpha}$
is constrained to satisfy Eq. (\ref{NotationsPureSpinorConstraint}).

Consider the subsheaf $U  \subset  TX$ consisting of vectors of the form
$u^{\alpha}{\partial\over\partial\theta^{\alpha}}$
(in other words, $\xi^{\alpha}=0$). Its space of sections is:
\begin{equation}
   \Gamma(U) = \{v\in \mbox{Vect}(X) \;|\; {\cal L}_v\lambda^{\alpha} = 0\}
   \end{equation}
We observe that $\Gamma(U)\subset \mbox{Vect}(X)$ is invariant under the action of $[Q^{(0)},\_]$.
Therefore, we can think of both $\Gamma(U)$ and $\Gamma(TX/U)$ as complexes with the differential $[Q^{(0)},\_]$.

\subsection{Summary of results for $H^1(Vect(X))$}\label{sec:SummaryOfH1}

Using the notations of Section \ref{IntroNotations}:
\begin{align} H^1(Vect(X))_{\rm odd}\;=\;{\bf C}\Big\langle
 &[(\!(\lambda\theta)\!)\otimes\theta]^{1/2}\otimes{\partial\over\partial\theta},\label{H1VectXOdd} \\  
 &(\!(\lambda\theta)\!)
        \left(
              \lambda^{\alpha}{\partial\over\partial\lambda^{\alpha}} +
              \theta^{\alpha}{\partial\over\partial\theta^{\alpha}}
              \right)
        \Big\rangle\nonumber{} \\ H^1(Vect(X))_{\rm even}\;=\;{\bf C}\Big\langle
 &[(\!(\lambda\theta)\!)\otimes\theta]^{1/2}\otimes
                                         \left(
                                               \lambda\Gamma_{mn}{\partial\over\partial\lambda} +
                                               \theta\Gamma_{mn}{\partial\over\partial\theta}
                                               \right)
                                         \Big\rangle\label{H1VectXEven} \end{align}
In the rest of this Section we will explain the computation.

\subsection{Exact sequences}\label{ExactSequences}

Consider the short exact sequence of complexes:

\begin{equation}
   0 \longrightarrow  \Gamma (U)  \longrightarrow  Vect(X) \longrightarrow  \Gamma(TX/U) \longrightarrow  0
   \end{equation}
The corresponding long exact sequence in cohomology of $[Q^{(0)},\_]$ is:

\begin{align} \longrightarrow\;
 &H^{n-1}(\Gamma(U))
              \longrightarrow
              H^{n-1}(Vect(X))
              \longrightarrow
              H^{n-1}(\Gamma(TX/U))
              \longrightarrow\nonumber{} \\ \longrightarrow\;
 &H^n(\Gamma(U))
              \longrightarrow
              H^n(Vect(X))
              \longrightarrow
              H^n(\Gamma(TX/U))
              \longrightarrow\nonumber{} \\ \longrightarrow\;
 &H^{n+1}(\Gamma(U))\longrightarrow
             \ldots\nonumber{} \end{align}

\subsection{Computation of $H^1(Vect(X))_{\rm odd}$}\label{sec:ComputationH1Odd}

\subsubsection{Summary of result}\label{sec:SummaryH1odd}

We use the following segment of the long exact sequence:
\begin{align}  
 &H^0(\Gamma(TX/U))_{\rm even}\stackrel{\delta}{\longrightarrow}\nonumber{} \\ \stackrel{\delta}{\longrightarrow}
 &H^1(\Gamma(U))_{\rm odd}
              \longrightarrow
              H^1(Vect(X))_{\rm odd}
              \longrightarrow
              H^1(\Gamma(TX/U))_{\rm odd}
              \stackrel{\delta}{\longrightarrow}\nonumber{} \\ \stackrel{\delta}{\longrightarrow}
 &H^2(\Gamma(U))_{\rm even}\nonumber{} \end{align}
The cohomology groups participating in this segment have the following description:
\begin{align} H^0(\Gamma(TX/U))_{even} \;=\;
 &\mbox{${\bf C}\langle D, M_{mn} \rangle$ of Eq. (\ref{DandM})}\nonumber{} \\ H^1(\Gamma(U))_{odd}\;=\;
 &{\bf C}\left\langle
                  (\lambda\Gamma^m\theta)(\theta\Gamma_m)_{\alpha}{\partial\over\partial\theta^{\beta}}
                  \right\rangle\label{AnswerH1GUodd} \\ \left[\delta\;:\;H^{0}(\Gamma(TX/U))_{even} \longrightarrow H^1(\Gamma(U))_{odd}\right]\;=\;
 &\mbox{$0$ Section \ref{sec:DeltaH0evenH1odd}}\label{AnswerDeltaH0evenH1odd} \\ H^1(\Gamma(TX/U))_{odd} \;=\;
 &{\bf C}
        \left\langle
        (\lambda\Gamma^m\theta)\left(\lambda^{\alpha}{\partial\over\partial\lambda^{\alpha}}\right)
        \right\rangle\label{AnswerGammaTXoverUodd} \\  
 &\mbox{Section \ref{H1DerLambda}}\nonumber{} \\ \left[\delta\;:\;H^{1}(\Gamma(TX/U))_{odd} \longrightarrow H^2(\Gamma(U))_{even}\right]\;=\;
 &0\label{AnswerDeltaH1oddH2even} \end{align}
This implies:
\begin{align} H^1(Vect(X))_{\rm odd}\;=\;
 &H^1(\Gamma(U))_{\rm odd} \oplus H^1(\Gamma(TX/U))_{\rm odd}\;=\;\nonumber{} \\ \;=\;
 &{\bf C}\left\langle\;
                  (\lambda\Gamma^m\theta)(\theta\Gamma_m)_{\alpha}{\partial\over\partial\theta^{\beta}}
                  \;,\;
                  (\lambda\Gamma^m\theta)
                  \left(
                        \lambda^{\alpha}{\partial\over\partial\lambda^{\alpha}} +
                        \theta^{\alpha}{\partial\over\partial\theta^{\alpha}}
                        \right)
                  \;\right\rangle\label{AnswerH1VectXodd} \end{align}

We will now explain the computation.

\subsubsection{Computation}\label{sec:ComputationH1VectOdd}

\paragraph{$\Gamma(TX/U)$}\label{sec:GammaTXU}

The space $\Gamma(TX/U)$ is generated as an ${\cal O}(X)$-module, by the following vector fields:
\begin{equation}
     D = \l ^\a  \frac{\p }{\p \l ^\a } \  ,\quad  M_{mn} = (\l \g _{mn})^\a \frac{\p }{\p \l ^\a }
     \label{DandM}\end{equation}
However $\Gamma(TX/U)$ is \emph{not} a free ${\cal O}(X)$ module, because there is a relation:
\begin{equation}
   \frac{1}{10}  (\l \g ^{mn})^\a  M_{mn} + \l ^\a  D = 0
   \end{equation}

\paragraph{$\delta\;:\; H^0(\Gamma(TX/U))_{\rm even} \longrightarrow H^1(\Gamma(U))_{\rm odd}$}\label{sec:DeltaH0evenH1odd}

It is zero because both $D$ and $M_{mn}$ can be extended to elements of $\mbox{Vect}(X)$ commuting with
$Q^{(0)}$:

\begin{align} D\mapsto 
 &\lambda^{\alpha}{\partial\over\partial\lambda^{\alpha}}
                   + \theta^{\alpha}{\partial\over\partial\theta^{\alpha}}\label{ExtensionOfD} \\ M_{mn} \mapsto
 &\left(\lambda\Gamma_{mn} {\partial\over\partial \lambda}\right) +
                \left(\theta\Gamma_{mn} {\partial\over\partial \theta}\right)\label{ExtensionOfM} \end{align}

\paragraph{$H^1(\Gamma(TX/U))_{odd}$ and $\delta \;:\; H^1(\Gamma(TX/U))_{odd} \rightarrow H^2(\Gamma(U))_{even}$}\label{H1DerLambda}

For any tensor $A^{lmn}$, consider vector fields of the form:
\begin{equation}
   A^{l,mn}
   (\lambda\Gamma_{l}\theta)
   \left(\lambda\Gamma_m\Gamma_n {\partial\over\partial\lambda}\right)
   \label{ElementOfC1GTXU}\end{equation}
Such vector fields generate $Z^1(\Gamma(TX/U))_{\rm odd}$.
But some of them are $Q^{(0)}$-exact:
\begin{equation}
     (\lambda\Gamma_{[l}\theta)
                    \left(\lambda \Gamma_{m]}\Gamma_n {\partial\over\partial\lambda}
                    \right) =
     {1\over 4}\left[
           Q^{(0)},
           (\theta\Gamma_{lm}\Gamma_p\theta)\left(\lambda \Gamma_p\Gamma_n {\partial\over\partial\lambda}\right)
           \right] \mbox{ \tt\small mod} \; \Gamma(U)
   \label{TrivialityOfC1GTXU}\end{equation}
Therefore the vector fields of the form Eq. (\ref{ElementOfC1GTXU}) with $A^{lmn}$ of the form:
\begin{equation}
     A^{l,mn} = X^{[lm]n} + Y^{l(mn)}\quad , \quad Y^{lmm} = 0
     \label{AZeroInH}\end{equation}
are zero in $H^1(\Gamma(TX/U))_{odd}$. This implies that $H^1(\Gamma(TX/U))_{odd}$ is generated by
the vector fields of the form:
\begin{equation}
     (\lambda\Gamma^i\theta)\left(\lambda^{\alpha}{\partial\over\partial\lambda^{\alpha}}\right)
   \label{LambdaThetaDilatation}\end{equation}
A vector field of Eq. (\ref{ElementOfC1GTXU}) is zero in cohomology iff:
\begin{equation}
   A^{l,lm} - A^{l,ml} - A^{m,ll} = 0
   \label{ProjectionOnH1GTX}\end{equation}
Vector fields of the form (\ref{LambdaThetaDilatation}) correspond to:
\begin{align}  
 &A^{l,mn} \;=\;{1\over 10}\delta_{il} \delta_{mn}\nonumber{} \\  
 &A^{l,lm} - A^{l,ml} - A^{m,ll} \;=\; -\delta_{im}\nonumber{} \end{align}
Notice that the section of $\Gamma(TX/U)$ defined by Eq. (\ref{LambdaThetaDilatation}) can be extended
to a $[Q^{(0)},\_]$-closed section of $TX$:
\begin{equation}
   (\lambda\Gamma^m\theta)
   \left(
         \left(\lambda^{\alpha}{\partial\over\partial\lambda^{\alpha}}\right) +
         \left(\theta^{\alpha}{\partial\over\partial\theta^{\alpha}}\right)
         \right)
   \end{equation}
This means that $\delta \;:\; H^1(\Gamma(TX/U))_{odd} \rightarrow H^2(\Gamma(U))_{even}$ is zero.

Eq. (\ref{TrivialityOfC1GTXU}) has the following refinement:

\begin{align}  
 &(\lambda\Gamma_{[l}\theta)
                       \left(
                             \lambda \Gamma_{m]}\Gamma_n {\partial\over\partial\lambda}
                          + \theta \Gamma_{m]}\Gamma_n {\partial\over\partial\theta}
                       \right) \;=\;\label{DownToU} \\ \;=\;
 &
           {1\over 4}\left[
                           Q^{(0)},
                           (\theta\Gamma_{lm}\Gamma_p\theta)
                           \left(
                                 \lambda \Gamma_p\Gamma_n {\partial\over\partial\lambda}
                                 +
                                 {1\over 3} \theta \Gamma_p\Gamma_n {\partial\over\partial\theta}
                                 \right)
                           \right]
           -
           {1\over 6} (\theta\Gamma_p\lambda)\left(\theta\Gamma_p\Gamma_{lm}\Gamma_n{\partial\over\partial\theta}\right)
           \nonumber{} \end{align}

\subsection{Computation of $H^1(\Gamma(TX))_{\rm even}$}\label{ComputationH1Even}

\subsubsection{Summary of result}\label{sec:SummaryH1even}

\begin{align} H^1(\mbox{Vect}(X))_{\rm even}\;=\;
 &{\bf C}\Big\langle
                  (\lambda\Gamma^m\theta)\Gamma_{m\alpha\beta}\theta^{\beta}
                  \left(
                        \lambda\Gamma_{kl}{\partial\over\partial\lambda}
                        + \theta\Gamma_{kl}{\partial\over\partial\theta}
                        \right)
                  \Big\rangle\label{AnswerH1VectXeven} \end{align}

\subsubsection{Computation}\label{sec:ComputationH1even}

We use the following segment of the long exact sequence:
\begin{align}  
 &H^0(\Gamma(TX/U))_{\rm odd}\stackrel{\delta}{\longrightarrow}\nonumber{} \\ \stackrel{\delta}{\longrightarrow}
 &H^1(\Gamma(U))_{\rm even}
              \longrightarrow
              H^1(Vect(X))_{\rm even}
              \longrightarrow
              H^1(\Gamma(TX/U))_{\rm even}
              \stackrel{\delta}{\longrightarrow}\nonumber{} \\ \stackrel{\delta}{\longrightarrow}
 &H^2(\Gamma(U))_{\rm odd}\nonumber{} \end{align}

\paragraph{$H^0(\Gamma(TX/U))_{odd}$}\label{H0DerLambdaOdd}
is generated by:
\begin{equation}
   Z_{\alpha}^q = (\Gamma_p\theta)_{\alpha}
   \left(\lambda\Gamma^p\Gamma^q{\partial\over\partial\lambda}\right)
\end{equation}

\paragraph{$H^1(\Gamma(U))_{even}$}\label{H1GUEven}
is generated by:
\begin{equation}
   Y^m_{\alpha} = (\theta\Gamma^m\lambda) {\partial\over\partial \theta^{\alpha}}
\end{equation}

\paragraph{Computation of $\delta \;:\; H^{0}(\Gamma(TX/U))_{odd} \longrightarrow H^1(\Gamma(U))_{even}$}\label{sec:CoboundaryH0OddH1Even}
\begin{align} \delta Z_{\alpha}^q\;=\;
 &(\Gamma_p\theta)_{\alpha}
                           \left(\lambda\Gamma^p\Gamma^q{\partial\over\partial\theta}\right)\;=\;\nonumber{} \\ \;=\;
 &- (\theta\Gamma^p\lambda)
           \left(\Gamma_p\Gamma^q{\partial\over\partial\theta}\right)_{\alpha} \mbox{ mod } [Q^{(0)},\_]
           = - (\Gamma_m\Gamma^q)_{\alpha}^{\beta} Y^m_{\beta} \mbox{ mod } [Q^{(0)},\_]\label{CoboundaryH0OddH1Even} \end{align}
The linear map $Y^q_{\alpha}\mapsto (\Gamma_m\Gamma^q)_{\alpha}^{\beta} Y^m_{\beta}$ is a bijection. More precisely:
\begin{equation}
   \left\{
         \lambda{\partial\over\partial\theta}\;,\;
         (\Gamma_p \theta)_{\alpha}\left(
                                         \theta\Gamma^p\Gamma^q{\partial\over\partial\theta} +
                                         2 \lambda\Gamma^p\Gamma^q{\partial\over\partial\lambda}
                                         \right)
         \right\} = - \left(\Gamma_p\Gamma^q \;(\theta\Gamma^p\lambda) {\partial\over\partial\theta}\right)_{\alpha}
         \end{equation}
\begin{align}  
 &
        \left\{
        \lambda{\partial\over\partial\theta}\;,\;
        \left(
              \delta^q_r\Gamma_p\theta - {1\over 10}\Gamma^q\Gamma_r\Gamma_p\theta
              \right)_{\alpha}
        \left(
              \theta\Gamma^p\Gamma^r{\partial\over\partial\theta} +
              2 \lambda\Gamma^p\Gamma^r{\partial\over\partial\lambda}
              \right)
        \right\} \; = 
        \nonumber{} \\ =\;
 &
           - 2 \left(
                     \left(
                           \delta^q_r - {1\over 10}\Gamma^q\Gamma_r
                           \right)           
                     \;(\theta\Gamma^r\lambda) {\partial\over\partial\theta}
                     \right)_{\alpha}
           \nonumber{} \end{align}
\begin{align}  
 &
        \left\{
        \lambda{\partial\over\partial\theta}\;,\;
        {1\over 10}(\Gamma^q\Gamma_r\Gamma_p \theta)_{\alpha}
        \left(
              \theta\Gamma^p\Gamma^r{\partial\over\partial\theta} +
              2 \lambda\Gamma^p\Gamma^r{\partial\over\partial\lambda}
              \right)
        \right\} \; = 
        \nonumber{} \\ =\;
 &
           8 \left(
                   {1\over 10}\Gamma^q\Gamma_r
                   \;(\theta\Gamma^r\lambda) {\partial\over\partial\theta}
                   \right)_{\alpha}
           \nonumber{} \end{align}
\begin{align} (\theta\Gamma^q\lambda){\partial\over\partial\theta^{\alpha}}\;=\;
 &\left\{ \lambda{\partial\over\partial\theta}
                   \;,\;
                   A^q_{\alpha}
                   \right\}
                   \nonumber{} \\  
 &\mbox{where}\nonumber{} \\ A^q_{\alpha}\;=\;
 &- {1\over 2} 
              \left(
                    \delta^q_r\Gamma_p\theta - {1\over 10}\Gamma^q\Gamma_r\Gamma_p\theta
                    \right)_{\alpha}
              \left(
                    \theta\Gamma^p\Gamma^r{\partial\over\partial\theta} +
                    2 \lambda\Gamma^p\Gamma^r{\partial\over\partial\lambda}
                    \right) \;+
              \nonumber{} \\  
 &
        + {1\over 80}(\Gamma^q\Gamma_r\Gamma_p \theta)_{\alpha}
        \left(
              \theta\Gamma^p\Gamma^r{\partial\over\partial\theta} +
              2 \lambda\Gamma^p\Gamma^r{\partial\over\partial\lambda}
              \right)\;=
        \nonumber{} \\ =\;
 &- {1\over 2} 
              \left(
                    \delta^q_r\Gamma_p\theta - {1\over 8}\Gamma^q\Gamma_r\Gamma_p\theta
                    \right)_{\alpha}
              \left(
                    \theta\Gamma^p\Gamma^r{\partial\over\partial\theta} +
                    2 \lambda\Gamma^p\Gamma^r{\partial\over\partial\lambda}
                    \right)
              \nonumber{} \end{align}

Therefore $H^{0}(\Gamma(TX/U))_{odd}$ cancels with $H^1(\Gamma(U))_{even}$.

\paragraph{Computation of $H^1(\Gamma(TX/U))_{\rm even}$ and vanishing of  $\delta \;:\; H^1(\Gamma(TX/U))_{even} \rightarrow H^2(\Gamma(U))_{odd}$}\label{sec:ComputationgH1Even}

\vspace{10pt}

The space of cocycles $Z^1(\Gamma(TX/U))_{\rm even}$ is generated by:
\begin{align}  
 &[(\!(\lambda\theta)\!)\otimes\theta]^{1/2} D\nonumber{} \\  
 &[(\!(\lambda\theta)\!)\otimes\theta]^{1/2} M_{mn}\nonumber{} \end{align}
where $D$ and $M_{mn}$ are from Eq. (\ref{DandM}).
Since both $D$ and $M_{mn}$ extend to $[Q^{(0)},\_]$-closed sections of $TX$ by Eqs. (\ref{ExtensionOfD})
and (\ref{ExtensionOfM}), the coboundary operator
$\delta \;:\; H^1(\Gamma(TX/U))_{even} \rightarrow H^2(\Gamma(U))_{odd}$ is zero.

But some cocycles are exact. Indeed, as sections of $TX/U$:

\begin{align} Q^{(0)}
 &\left(
                 \Gamma_{pq}\Gamma_m\theta\,
                 (\theta\Gamma_{kpq}\theta)
                 \left(\lambda \Gamma^k\Gamma^m {\partial\over\partial\lambda}\right)
                 \right)
                \;=\;\nonumber{} \\ \;=\;
 &-32 \Gamma_m\theta(\theta\Gamma^m\lambda)
               \left(\lambda{\partial\over\partial\lambda}\right)
               -
               4\Gamma_n \Gamma_m \Gamma_q \theta (\theta \Gamma^q\lambda)
               \left(\lambda\Gamma^n\Gamma^m {\partial\over\partial\lambda}\right)\nonumber{} \end{align}

\section{Coefficients of normal form satisfy wave equations}\label{WaveEquations}

Modulo $F^4\mbox{Vect}$ we can choose the coordinates so that:
\begin{align} Q_L\;=\;
 &\lambda_L{\partial\over\partial\theta_L}
                    + (\!(\lambda_L\theta_L)\!)^mE^{Ln}_m{\partial\over\partial x^n} \;+\label{LeadingTermsOfQL1} \\  
 &+ \; (\!(\lambda_L\theta_L)\!)^m
          \left(
                \lambda_L \omegaLLL_m \partial_{\lambda_L} + \theta_L \omegaLLL_m \partial_{\theta_L} +
                \lambda_R \omegaLRR_m \partial_{\lambda_R} + \theta_R \omegaLRR_m \partial_{\theta_R} 
                \right)
          \;  +\label{LeadingTermsOfQL2} \\  
 &+ \; (\!(\lambda_L\theta_L\theta_L)\!)
          \left(P_{LL}{\partial\over\partial \theta_L} + P_{LR} {\partial\over\partial \theta_R}\right)
          \mbox{ mod }F^4\label{LeadingTermsOfQL} \\ Q_R\;=\;
 &\lambda_R{\partial\over\partial\theta_R} +
                    (\!(\lambda_R\theta_R)\!)^mE^{Rn}_m{\partial\over\partial x^{n}}\;+\nonumber{} \\  
 &+ \; (\!(\lambda_R\theta_R)\!)^m
          \left(
                \lambda_R \omegaRRR_m \partial_{\lambda_R} + \theta_R \omegaRRR_m \partial_{\theta_R} +
                \lambda_L \omegaRLL_m  \partial_{\lambda_L} + \theta_L \omegaRLL_m \partial_{\theta_L} 
                \right)
          \;  +\nonumber{} \\  
 &+ (\!(\lambda_R\theta_R\theta_R)\!)
          \left(P_{RL}{\partial\over\partial \theta_L} + P_{RR} {\partial\over\partial \theta_R}\right)
          \mbox{ mod } F^4\label{LeadingTermsOfQR} \end{align}
where $E,\Omega,P$ are some functions of $x$.
Indeed, using
Section \ref{sec:FirstPage}:
\begin{itemize}\item $H^1(C^{\rm fun}_L)_{\rm odd}\otimes {\partial\over\partial x}$
          enters on Line (\ref{LeadingTermsOfQL1}),\item Second part of $H^1(C^{\rm vect}_L)_{\rm odd}$ (see Eq. (\ref{AnswerH1VectXodd}))
          and $H^1(C^{\rm fun}_L)_{\rm odd}\otimes H^0(C_R^{\rm vect})_{\rm even}$
          on Line (\ref{LeadingTermsOfQL2}),\item First part of $H^1(C_L^{\rm vect})_{\rm odd}$ (see Eq. (\ref{AnswerH1VectXodd}))
          and $H^1(C_L^{\rm fun})_{\rm even}\otimes H^0(C_R^{\rm vect})_{\rm odd}$
          on Line (\ref{LeadingTermsOfQL})\end{itemize}

\subsection{Equations for tetrad and spin connection}\label{sec:EquationsForTetradAndSpinConnections}

\subsubsection{Fixing  $(so(10)\oplus {\bf C})_L$ and $(so(10)\oplus {\bf C})_R$}\label{sec:FixingFrame}

Let us study the linearized order in deviations from flat space-time. In flat space-time
$E^{L\mu}_m = E^{R\mu}_m = \delta^{\mu}_m$. The deviation from
flatness can be written as:
\begin{equation}
   E^{L\mu}_m = \delta^{\mu}_m + \delta^{\mu}_n e^{L}_{n,m}
   \quad \mbox{\tt\small and} \quad
   E^{R\mu}_m = \delta^{\mu}_m + \delta^{\mu}_n e^{R}_{n,m}
   \end{equation}
where $e^L$ and $e^R$ are infinitesimal. We assume summation over repeated indices.
We can choose a freedom of $so(10)\oplus {\bf C}$ redefinitions
of both $(\lambda_L,\theta_L)$ and $(\lambda_R,\theta_R)$ to fix:
\begin{align}  
 &e^L_{[m,n]} = e^R_{[m,n]} \;=\; 0\label{BothEAreSymmetric} \\  
 &e^L_{m,m} = e^R_{m,m}\label{TraceELisTraceER} \end{align}
At this point, the only remaining freedom in redefinition of $\lambda$ and $\theta$
is overall rescaling of $(\lambda_L,\lambda_R, \theta_L, \theta_R)$.
We fixed both $(so(10)\oplus {\bf C})_L \oplus (so(10)\oplus {\bf C})_R$
down to the diagonal $\bf C$.

\subsubsection{Fixing $\omegaLLL$ and $\omegaRRR$}\label{sec:FixingOmegaLLLandOmegaRRR}

According to Section \ref{H1DerLambda}, we can choose:
\begin{equation}
     \omegaLLL_{m,lk} = {1\over 10}\omegaLLL^{(s)}_m\delta_{lk}
     \label{FixingOmegaLLL}\end{equation}
From $\{Q_L,Q_R\} = 0$, the coefficient of
$(\!(\lambda_L\theta_L)\!)^l(\!(\lambda_R\theta_R)\!)^n
                                           \left(
                                                 \lambda_L{\partial\over\partial\lambda_L}
                                                 + \theta_L{\partial\over\partial\theta_L}
                                                 \right)
$, projected to $H^1(\Gamma(TX/U))$ (see Eq. (\ref{ProjectionOnH1GTX})):
\begin{equation}
     2\partial_m\omegaRLL_{n,[ml]} - \partial_l\omegaRLL_{n,mm} + \partial_n\omegaLLL^{(s)}_l = 0
     \label{OmegaLLLvsOmegaRLL}\end{equation}
Similarly with $L\leftrightarrow R$:
\begin{equation}
     2\partial_m\omegaLRR_{n,[ml]} - \partial_l\omegaLRR_{n,mm} + \partial_n\omegaRRR^{(s)}_l = 0
     \label{OmegaRRRvsOmegaLRR}\end{equation}
Eqs. (\ref{FixingOmegaLLL}) and (\ref{OmegaLLLvsOmegaRLL})
and similar equations with $L\leftrightarrow R$
determine $\omegaLLL$ and $\omegaRRR$ in terms of $\omegaRLL$ and $\omegaLRR$.
Eqs. (\ref{OmegaLLLvsOmegaRLL}) and (\ref{OmegaRRRvsOmegaLRR}) guarantee the cancellation of the obstacle
in $H^1(\Gamma(TX/U))_{\rm odd}$, \textit{i.e.} the one containing $\partial\over\partial\lambda$.
It remains the coefficient of $(\lambda\Gamma^m\theta)(\theta\Gamma_m)_{\alpha}{\partial\over\partial\theta^{\beta}}$
(see Eq. (\ref{AnswerH1VectXodd})).
This will cancel by $P_{LL}$ and $P_{RR}$, see Section \ref{EqsRRQLQR}.

Let us denote:
\begin{equation}
   \Omega^L_{m,nk} = - \omegaRLL_{m,[nk]} + {1\over 2}\omegaRLL_{m,pp}\delta_{nk}
   \end{equation}
and similar definition for $\Omega^R_{m,nk}$ in terms of $\omegaLRR$.

This notation is useful, because for any vector $V_l$:
\begin{equation}
   \omegaRLL_{m,nk}V_l\left(\Gamma_n\Gamma_k\Gamma_l + \Gamma_l (\Gamma_n\Gamma_k)^T\right)
   \;=\;
   4V_p \Omega^L_{m,pq} \Gamma_q
   \end{equation}
From  $\{Q_L,Q_R\} = 0$, the coefficient of $(\!(\lambda_L\theta_L)\!)^m(\!(\lambda_R\theta_R)\!)^n{\partial\over\partial x}$:
\begin{align} {\partial\over\partial x^m}E^R_n + E^R_k\Omega^R_{m,kn} \;=\;
 &{\partial\over\partial x^n}E^L_m + E^L_k\Omega^L_{n,km}\nonumber{} \\ \partial_m e_{Rn,k} + \Omega^R_{m,nk} \;=\;
 &\partial_n e_{Lm,k} + \Omega^L_{n,mk}\nonumber{} \end{align}
Equivalently:
\begin{align}  
 &\partial_{[m}(e_L + e_R)_{n],k} + \left(\Omega^L + \Omega^R\right)_{[m,n]k} = 0\label{TorsionZero} \\  
 &\partial_{(m}(e_L - e_R)_{n),k} + \left(\Omega^L - \Omega^R\right){}_{(m,n)k} = 0\label{AntiTorsionZero} \end{align}
Eq. (\ref{TorsionZero}) is zero torsion of the ``average'' (\textit{i.e.} left plus right)
connection.

\subsubsection{Einstein equations}\label{sec:EinsteinEquations}

Let us denote:
\begin{align} g_{mn} \;=\;
 &e_{L(m,n)} + e_{R(m,n)}\nonumber{} \\ \Omega_{m,nk} \;=\;
 &{1\over 2}\left(\Omega^L + \Omega^R\right)_{m,nk}\nonumber{} \end{align}
Then Eq. (\ref{TorsionZero}) implies the existence of $a_m$ such that:
\begin{equation}
     \Omega_{m,nk} = - g_{m[n}\stackrel{\leftarrow}{\partial}_{k]} + a_m \delta_{nk} - 2 \delta_{m[n}a_{k]}
     \label{DefAm}\end{equation}
Infinitesimal coordinate redefinition $\tilde{x}^{\mu} = x^{\mu} + \varepsilon v^{\mu}$, followed a compensating rotation
of $\theta$ and $\lambda$ in order to preserve Eq. (\ref{BothEAreSymmetric}), corresponds to:

\begin{align} \delta_v e_{Lm,k} \;=\;
 &\delta_v e_{Rm,k} = \partial_{(m} v_{k)}\nonumber{} \\ \delta_v \Omega_{m,nk} \;=\;
 &\partial_m(\partial_{[n}v_{k]})\nonumber{} \\ \delta\omegaLLL^{(s)}_m \;=\;
 &2\partial_p \partial_{[p}v_{n]}\nonumber{} \end{align}

The overall rescaling
\begin{equation}
   \delta_{\gamma} (\lambda_L,\lambda_R, \theta_L, \theta_R) =
   (\gamma\lambda_L,\gamma\lambda_R, \gamma\theta_L, \gamma\theta_R)
   \label{rescalingGaugeTransformation}\end{equation}
corresponds to:
\begin{align} \delta_{\gamma} g_{m,n} \;=\;
 &2\gamma\delta_{mn}\label{RescalingG} \\ \delta_{\gamma} a_m \;=\;
 &-\partial_m\gamma\label{RescalingA} \\ \delta_{\gamma} \Omega_{m,nk} \;=\;
 &-\partial_m\gamma \delta_{nk}\label{RescalingO} \end{align}
From $\{Q_L,Q_L\} = 0$ and $\{Q_R,Q_R\} = 0$ follows that $\omegaLRR_m$ and $\omegaRLL_m$ both satisfy
Maxwell equations:

\begin{equation}
     {\partial\over\partial x^n}{\partial\over\partial x^{[m}} \omegaLRR_{n]pq} =
     {\partial\over\partial x^n}{\partial\over\partial x^{[m}} \omegaRLL_{n]pq} = 0
     \label{MaxwellEquations}\end{equation}
Considering the scalar part, we conclude that $a_m$ satisfies the Maxwell equations:
\begin{equation}
   \partial_m \partial_{[m}a_{n]} = 0
   \end{equation}
and $g_{mn}$ satisfies:

\begin{align}  
 &\partial_p\partial_{[p}g_{m][n}\stackrel{\leftarrow}{\partial}_{k]} + 2\partial_p \partial_{[p}\delta_{m][n}a_{k]} = 0\nonumber{} \\ \Rightarrow
 &\partial_k \left(
                            2\partial_{[p}g_{n][m}\stackrel{\leftarrow}{\partial}_{p]} +
                            \partial_m a_n + \delta_{mn} \partial^pa_p
                            \right)
                      - (k\leftrightarrow n) = 0\label{CurlRicci} \\ \Rightarrow
 &\exists b_m\;:\;
           2\partial_{[p}g_{n][m}\stackrel{\leftarrow}{\partial}_{p]} +
           \partial_m a_n +
           \delta_{mn} \partial^pa_p  
           = - \partial_n b_m\nonumber{} \end{align}
It follows from the symmetry $m\leftrightarrow n$ that exists $\phi$ such that
$b_m  = a_m - \partial_m\phi$. Therefore:
\begin{equation}
     2\partial_{[p}g_{n][m}\stackrel{\leftarrow}{\partial}_{p]} +
     \delta_{mn} \partial^p a_p +
     2\partial_{(m}a_{n)} = \partial_m\partial_n\phi
     \label{EqWithAandPhi}\end{equation}
The rescaling Eqs. (\ref{RescalingG}), (\ref{RescalingA}) and (\ref{RescalingO}) are accompanied by:
\begin{equation}
   \delta_{\gamma} \phi = (10-4)\gamma
   \end{equation}

With Eq. (\ref{CurlRicci}), the consistency of the sum of Eq. (\ref{OmegaLLLvsOmegaRLL}) and Eq. (\ref{OmegaRRRvsOmegaLRR})
requires, modulo zero modes:
\begin{equation}
   \partial_{[m}a_{n]} = 0
   \end{equation}
and therefore $a_m$ can be gauged away:
\begin{equation}
   a_m = 0
   \end{equation}
fixing the overall rescaling gauge symmetry of Eq. (\ref{rescalingGaugeTransformation}).

\subsubsection{Antisymmetric tensor}\label{sec:AntisymmetricTensor}

Eq. (\ref{AntiTorsionZero}) implies, after total symmetrization:
\begin{equation}
   \partial_{(m}(e_L - e_R)_{n,k)} + \Omega^{L(s)}_{(m}\delta_{nk)} - \Omega^{R(s)}_{(m}\delta_{nk)} = 0
   \label{SymmetrizedDEm}\end{equation}
Modulo finite dimensional spaces, Eqs. (\ref{SymmetrizedDEm}), (\ref{AntiTorsionZero}) and (\ref{TraceELisTraceER})  imply that (\textit{cf} Eq. (\ref{GeneralEquationModuloDelta})):
\begin{align} e_L - e_R \;=\;
 &0\nonumber{} \\ \Omega^{L(s)}_m - \Omega^{R(s)}_m \;=\;
 &0\label{DifferenceOfScalarConnections} \\ \left(\Omega^L - \Omega^R\right){}_{(m,n)k} \;=\;
 &0\nonumber{} \end{align}
Therefore $\Omega^L - \Omega^R$ is antisymmetric:
\begin{equation}
   \left(\Omega^L - \Omega^R\right)_{k,lm} = H_{klm} = H_{[klm]}
   \end{equation}
Eqs. (\ref{MaxwellEquations}) imply:
\begin{equation}
   \partial^p\partial_{[p}H_{q]mn} = 0
   \label{MaxwellForH}\end{equation}

The consistency of the difference of Eq. (\ref{OmegaLLLvsOmegaRLL}) and Eq. (\ref{OmegaRRRvsOmegaLRR})
implies  that $H_{lmn}$ is harmonic:
\begin{equation}
   \partial_p\partial^p H_{lmn} = 0
   \end{equation}
and, modulo a constant, divergenceless:
\begin{equation}
   \partial^p H_{pmn} = 0
   \label{DivHisZero}\end{equation}
Then:

\begin{equation}
   \omegaLLL_l^{(s)} = \omegaRRR_l^{(s)} = 2 \partial_l \phi - 2 \partial_{[l}g_{m]m}
   \end{equation}

\subsection{Equations for bispinors}\label{sec:EquationsForRR}

\subsubsection{Equations following from $\{Q_L,Q_R\}=0$}\label{EqsRRQLQR}

Considering terms proportional to $(\!(\lambda_R\theta_R)\!)(\!(\lambda_L\theta_L\theta_L)\!){\partial\over\partial\theta_L}$
and similar terms with $L\leftrightarrow R$, we need to require that they cancel similar terms in
Section \ref{sec:FixingOmegaLLLandOmegaRRR}.

\begin{align}  
 &
      {\partial\over\partial x^m} P_{RR}^{\hat{\alpha}\hat{\beta}} =
      {1\over 6}\partial_{[p} H_{qr]m}\Gamma_{pqr}^{\hat{\alpha}\hat{\beta}}
      + {2\over 3}\partial_m(\partial_l\phi - \partial_{[l}g_{p]p})\Gamma_l^{\hat{\alpha}\hat{\beta}}
      \nonumber{} \\  
 &
      {\partial\over\partial x^m} P_{LL}^{\alpha\beta} =
      - {1\over 6}\partial_{[p} H_{qr]m}\Gamma_{pqr}^{\alpha\beta}
      + {2\over 3}\partial_m(\partial_l\phi - \partial_{[l}g_{p]p})\Gamma_l^{\alpha\beta}
      \nonumber{} \end{align}
This implies that modulo zero modes:
\begin{align}  
 &\partial_{[k}H_{lmn]} = 0\nonumber{} \\  
 &
        P_{RR}^{\hat{\alpha}\hat{\beta}} =
        {1\over 18}H_{klm}\Gamma_{klm}^{\hat{\alpha}\hat{\beta}}
        + {2\over 3}(\partial_l\phi - \partial_{[l}g_{p]p})\Gamma_l^{\hat{\alpha}\hat{\beta}}
        \nonumber{} \\  
 &
        P_{LL}^{\alpha\beta} =
        - {1\over 18}H_{klm}\Gamma_{klm}^{\alpha\beta}
        + {2\over 3}(\partial_l\phi - \partial_{[l}g_{p]p})\Gamma_l^{\alpha\beta}        
        \nonumber{} \end{align}
The antisymmetric tensor field $H_{lmn}$ should be identified with the field strength of
the NSNS B-field: $H = dB$. 

Now consider the terms proportional to $(\!(\lambda_R\theta_R\theta_R)\!)(\!(\lambda_L\theta_L)\!){\partial\over\partial\theta_L}$:
\begin{equation}
   (\!(\lambda_R\theta_R\theta_R)\!)_{\hat{\alpha}}\;
   \partial_m P_{RL}^{\hat{\alpha}\alpha}\;
   (\!(\lambda_L\theta_L)\!)^m{\partial\over\partial\theta_L^{\alpha}}
   \end{equation}
It is cancelled by adding:
\begin{equation}
   -{1\over 2}
   (\!(\lambda_R\theta_R\theta_R)\!)_{\hat{\alpha}}\;
   \partial_m P_{RL}^{\hat{\alpha}\alpha}\;
   \left(
         \delta^m_r\Gamma_p\theta - {1\over 8}\Gamma^m\Gamma_r\Gamma_p\theta
         \right)_{\alpha}
   \left(
         \theta_L\Gamma^p\Gamma^r{\partial\over\partial\theta_L} +
         2 \lambda_L\Gamma^p\Gamma^r{\partial\over\partial\lambda_L}
         \right)
   \end{equation}
leading to the extra term:
\begin{equation}
   {1\over 3}
   (\!(\lambda_R\theta_R\theta_R)\!)_{\hat{\alpha}}\;
   \partial_m P_{RL}^{\hat{\alpha}\alpha}\;
   \left((4\delta_{mq} - \Gamma_m\Gamma_q)(\!(\lambda_L\theta_L\theta_L)\!)\right)_{\alpha}\;
   {\partial\over\partial x^q}
   \end{equation}
There is a similar contribution with $R\leftrightarrow L$. For them to cancel each other, we need:
\begin{align}  
 &\Gamma^m_{\alpha\beta}\partial_m P_{RL}^{\hat{\alpha}\beta} = 0\nonumber{} \\  
 &\Gamma^m_{\hat{\alpha}\hat{\beta}}\partial_m P_{LR}^{\alpha\hat{\beta}} = 0\nonumber{} \\  
 &P_{LR}^{\alpha\hat{\alpha}} = P_{RL}^{\hat{\alpha}\alpha}\nonumber{} \end{align}

\subsubsection{Equations for RR bispinor following from $\{Q_L,Q_L\}=0$}\label{EqsRRQLQL}

To get $\{Q_L, Q_L\} = 0$ and $\{Q_R, Q_R\} = 0$ we need:
\begin{align}  
 &\Gamma^m_{\beta\alpha} {\partial\over\partial x^m} P_{LR}^{\alpha\hat{\alpha}} = 0\nonumber{} \\  
 &\Gamma^m_{\hat{\beta}\hat{\alpha}} {\partial\over\partial x^m} P_{RL}^{\alpha\hat{\alpha}} = 0\nonumber{} \end{align}

\section{Fermionic fields}\label{FermionicFields}

In Section \ref{WaveEquations} we restricted ourselves with $Q_L$ and $Q_R$
parameterized by \emph{even} functions $E^{L},E^{R},\ldots$.  We will now add the terms
parameterized by \emph{odd} functions.
According to Section \ref{ComputationH1Even} these terms are:
\begin{align} Q_L'\;=\;
 &\xi_{LRm}^{\hat{\beta}}(x)
            (\!(\lambda_L\theta_L)\!)^m {\partial\over\partial\theta_R^{\hat{\beta}}}\;+
            (\!(\lambda_L\theta_L\theta_L)\!)_{\alpha} \psi_L^{\alpha\mu}(x) {\partial\over\partial x^{\mu}}\;+
            \nonumber{} \\  
 &+\;\xiLLL^{\alpha [mn]}(x)(\!(\lambda_L\theta_L\theta_L)\!)_{\alpha}
    \left(
                       \lambda_L\Gamma_{mn}{\partial\over\partial\lambda_L}
                       + \theta_L\Gamma_{mn}{\partial\over\partial\theta_L}
                       \right)     \;+\nonumber{} \\  
 &+\;\xiLRR^{\alpha mn}(x)(\!(\lambda_L\theta_L\theta_L)\!)_{\alpha}
    \left( 
                       \lambda_R\Gamma_m\Gamma_n{\partial\over\partial\lambda_R}
                       + \theta_R\Gamma_m\Gamma_n{\partial\over\partial\theta_R}
                       \right)    \;+\ldots\nonumber{} \end{align}
\begin{align} Q_R'\;=\;
 &\xi_{RLm}^{\beta}(x)
            (\!(\lambda_R\theta_R)\!)^m {\partial\over\partial\theta_L^{\beta}}\;+
            (\!(\lambda_R\theta_R\theta_R)\!)_{\hat{\alpha}} \psi_R^{\hat{\alpha}\mu}(x) {\partial\over\partial x^{\mu}}\;+
            \nonumber{} \\  
 &+\;\xiRRR^{\hat{\alpha} [mn]}(x)(\!(\lambda_R\theta_R\theta_R)\!)_{\hat{\alpha}}
    \left(
                       \lambda_R\Gamma_{mn}{\partial\over\partial\lambda_R}
                       + \theta_R\Gamma_{mn}{\partial\over\partial\theta_R}
                       \right)     \;+\nonumber{} \\  
 &+\;\xiRLL^{\hat{\alpha} mn}(x)(\!(\lambda_R\theta_R\theta_R)\!)_{\hat{\alpha}}
    \left( 
                       \lambda_L\Gamma_m\Gamma_n{\partial\over\partial\lambda_L}
                       + \theta_L\Gamma_m\Gamma_n{\partial\over\partial\theta_L}
                       \right)    \;+\ldots\nonumber{} \end{align}
The first terms in both $Q_L'$ and $Q_R'$ are of grade 1, and the rest of the terms are of grade 3.

\subsection{Grade 3 terms are determined by the grade 1 terms}\label{sec:Grade3AreFixed}

Let us first assume that the grade 1 terms are zero, \textit{i.e} $\xi_{LRm}^{\hat{\beta}}(x)=0$
and $\xi_{RLm}^{\beta}(x)=0$.
Considering the coefficient of $(\!(\lambda_L\theta_L\theta_L)\!)(\!(\lambda_R\theta_R)\!){\partial\over\partial x}$, we deduce that $\psi_L^{\alpha\mu}$ satisfies:
\begin{equation}
   \partial^{\nu}\psi_L^{\alpha\mu}
   +
   4\xiLRR^{\alpha [\nu\mu]}
   -
   2\xiLRR^{\alpha mm} g^{\mu\nu} 
   = 0
   \end{equation}
and a similar equation for $\psi_R^{\hat{\alpha}\mu}$. This implies
(see  Section \ref{FiniteDimensionalSolutions})
that modulo finite dimensional subspaces (which we ignore):
\begin{align} \psi_L^{\alpha\mu} \;=\;
 &0\nonumber{} \\ \xiLRR^{\alpha \nu\mu} \;=\;
 &0\nonumber{} \\ \psi_R^{\hat{\alpha}\mu} \;=\;
 &0\nonumber{} \\ \xiRLL^{\hat{\alpha} \nu\mu} \;=\;
 &0\nonumber{} \end{align}
The coefficients $\xi^{\hat{\alpha}}_{LRm}$ and $\xi_{RLm}^{\alpha}$ come with gauge transformations:
\begin{align} \delta_{\phi_L}\xi_{LRm}^{\hat{\alpha}}\;=\;
 &\partial_m \phi_L^{\hat{\alpha}}\nonumber{} \\ \delta_{\phi_R}\xi_{RLm}^{\alpha}\;=\;
 &\partial_m \phi_R^{\alpha}\nonumber{} \end{align}
Considering the coefficient of
$(\!(\lambda_L\theta_L\theta_L)\!)(\!(\lambda_R\theta_R)\!)
                              \left(\lambda_L{\partial\over\partial\lambda_L} + \theta_L{\partial\over\partial\theta_L}\right)$,
we conclude that $\xiLLL^{\alpha [mn]}$
(and similarly $\xiRRR^{\hat{\alpha} [mn]}$)
are constants, and we ignore them.

\subsection{Grade 1 terms}\label{sec:Grade1Terms}

\subsubsection{Requiring $Q_L^2 = 0$}\label{sec:FermionicQLQL}

Requiring $Q_L^2 = 0$, the ``Maxwell bishop move'':
\begin{align}  
 &\xi_{LRm}^{\hat{\beta}}(x)(\!(\lambda_L\theta_L)\!)^m
            \stackrel{(\!(\lambda_L\theta_L)\!)\partial_x}{\longrightarrow}
            \partial_n\xi_{LRm}^{\hat{\beta}}(x)(\!(\lambda_L\theta_L)\!)^m(\!(\lambda_L\theta_L)\!)^n
            \stackrel{(\lambda_L\partial_{\theta_L})^{-1}}{\longrightarrow}
            \ldots
            \stackrel{(\!(\lambda_L\theta_L)\!)\partial_x}{\longrightarrow}\nonumber{} \\ \longrightarrow
 &\partial^n\partial_{[n}\xi_{LRm]}^{\hat{\beta}}(x)(\!(\lambda_L\theta_L)\!)^p(\!(\lambda_L\theta_L)\!)^q(\!(\theta_L\theta_L)\!)^{pqm}\nonumber{} \end{align}
we conclude that $\xi_{LR}$ (and similarly $\xi_{RL}$) should satisfy the Maxwell equations:
\begin{equation}
   \partial_m \partial_{[m}\xi_{LRn]} = 0
   \end{equation}

\subsubsection{Requiring $\{Q_L,Q_R\} = 0$}\label{sec:FermionicQLQR}

We will now consider the anticommutator of $Q_R'$ with $Q_L$. It is convenient to
start by completing $\partial\over\partial\theta_L^{\beta}$ to
${\partial\over\partial\theta_L^{\beta}} - (\Gamma^m\theta_L)_{\beta}{\partial\over\partial x^m}$.
We have:
\begin{align}  
 &
        \left\{\;
        \lambda_L{\partial\over\partial\theta_L} + (\lambda_L\Gamma^m\theta_L){\partial\over\partial x^m}
        \;,\;
        (\lambda_R\Gamma^n\theta_R)
        \xi_{RLn}^{\beta}\left(
                                {\partial\over\partial\theta_L^{\beta}} -
                                (\Gamma^k\theta_L)_{\beta}{\partial\over\partial x^k}
                                \right)
        \;\right\}\;=
        \nonumber{} \\ \;=\;
 &(\lambda_R\Gamma^n\theta_R)
           {\partial\xi_{RLn}^{\beta} \over\partial x^m}
           (\lambda_L\Gamma^m\theta_L)
           \left(
                 {\partial\over\partial\theta_L^{\beta}} -
                 (\Gamma^k\theta_L)_{\beta}{\partial\over\partial x^k}
                 \right)\nonumber{} \end{align}
The term $(\lambda_R\Gamma^n\theta_R){\partial\xi_{RLn}^{\beta} \over\partial x^m}(\lambda_L\Gamma^m\theta_L){\partial\over\partial\theta_L^{\beta}}$ is removed by further modifying (\textit{cp} Eq. (\ref{CoboundaryH0OddH1Even})):
\begin{equation}
   (\theta_R\Gamma^k\lambda_R)\;\xi_{RLk}^{\beta}
       \left[
             {\partial\over\partial\theta_L^{\beta}} - (\Gamma^n\theta_L)_{\beta}{\partial\over\partial x^n}
             \right]
       \end{equation}

to:
\begin{align}  
 &(\theta_R\Gamma^k\lambda_R)\xi_{RLk}^{\beta}
                                   \left[
                                         {\partial\over\partial\theta_L^{\beta}} -
                                         (\Gamma^n\theta_L)_{\beta}{\partial\over\partial x^n}
                                         \right]\;+\nonumber{} \\  
 &+ {1\over 2}(\theta_R\Gamma^k\lambda_R){\partial\xi_{RLk}^{\alpha}\over\partial x^m}
          \left(
                \left(\delta_{mn} - {1\over 8}\Gamma_m\Gamma_n\right)
                \Gamma_p\theta_L
                \right)_{\alpha}
          \left[
                2\lambda_L \Gamma_p\Gamma_n {\partial\over\partial\lambda_L}
                +\theta_L  \Gamma_p\Gamma_n {\partial\over\partial\theta_L}
                \right]\nonumber{} \end{align}
Then, when we commute with $(\lambda_l\Gamma^m\theta_L){\partial\over\partial x^m}$, this modification produces:
\begin{equation}
   {1\over 2}(\theta_R\Gamma^k\lambda_R){\partial\xi_{RLk}^{\alpha}\over\partial x^m}
   \left(
         \left(\delta_{mn} - {1\over 8}\Gamma_m\Gamma_n\right)
         \Gamma_p\theta_L
         \right)_{\alpha}
   (2(\lambda_L \Gamma_p\Gamma_n \Gamma_q\theta_L) + (\theta_L \Gamma_p\Gamma_n \Gamma_q\lambda_L))
   {\partial\over\partial x^q}
   \end{equation}
which should cancel  $-(\lambda_R\Gamma^k\theta_R){\partial\xi_{RLk}^{\alpha} \over\partial x^m}(\lambda_L\Gamma^m\theta_L)(\Gamma^q\theta_L)_{\alpha}{\partial\over\partial x^q}$. Indeed, let us denote:
\begin{equation}
   A^{mq}_{\alpha} \;=\;
   - (\lambda_L\Gamma^m\theta_L)(\Gamma^q\theta_L)_{\alpha}
   \; + \;
   {1\over 2}\left(
         \left(\delta_{mn} - {1\over 8}\Gamma_m\Gamma_n\right)
         \Gamma_p\theta_L
         \right)_{\alpha}
   \left(2 (\lambda_L \Gamma_p\Gamma_n \Gamma_q\theta_L) + (\theta_L \Gamma_p\Gamma_n \Gamma_q\lambda_L)\right)
   \end{equation}

The total contribution is:
\begin{equation}
   (\theta_R\Gamma^k\lambda_R){\partial\xi_{RLk}^{\alpha}\over\partial x^m}A_{\alpha}^{mq}{\partial\over\partial x^q}
   \label{TotalContributionToCancel}\end{equation}
We observe:
\begin{equation}
   \lambda_L{\partial\over\partial\theta_L} A^{mq}_{\alpha} = 0
   \end{equation}

In fact:
\begin{equation}
   A^{mq}_{\alpha} \;=\; {1\over 3} \left(\left( 4\delta_{mq} - \Gamma_m \Gamma_q \right)\Gamma_p\theta_L\right)_{\alpha}
   (\lambda_L\Gamma_p\theta_L) + \lambda_L{\partial\over\partial\theta_L}(\ldots)
   \end{equation}
To cancel Eq. (\ref{TotalContributionToCancel}) we must impose the Dirac equation on $\xi^{\alpha}_{RLk}$, in the following sense. 
Require that exists $\eta_{RL\alpha}$ such that:
\begin{equation}
   \Gamma^p_{\alpha\beta}{\partial\over\partial x^p}\xi_{RLk}^{\beta} = {\partial\over\partial x^k} \eta_{RL\alpha}
   \end{equation}
Then we cancel Eq. (\ref{TotalContributionToCancel}) by choosing
\begin{align} \psi_L^{\alpha q} \;=\;
 &{1\over 3}\Gamma_q^{\alpha\beta} \eta_{RL\beta} + {4\over 3}\xi^{\alpha}_{RLq}\nonumber{} \\ \xiLRR^{\alpha mq}(x) \;=\;
 &{4\over 3}\partial_{[m}\xi_{RLq]}^{\alpha}\nonumber{} \end{align}

To summarize:
\begin{align} Q_L' \;=\;
 &(\!(\lambda_L\theta_L)\!)^m\xi_{LRm}^{\hat{\beta}}(x)
                                    \left(
                                          {\partial\over\partial\theta_R^{\hat{\beta}}}
                                          -
                                          (\Gamma^n\theta_R)_{\hat{\beta}}
                                          {\partial\over\partial x^n}
                                          \right)\;+
                                    \nonumber{} \\  
 &+\;(\!(\lambda_L\theta_L)\!)^k{\partial \xi_{LRk}^{\hat{\beta}}\over\partial x^m}
                             \left(
                                   \left(\delta_{mn} - {1\over 8}\Gamma_m\Gamma_n\right)
                                   \Gamma_p\theta_R
                                   \right)_{\alpha}
                             \left[
                                   2\lambda_R \Gamma_p\Gamma_n {\partial\over\partial\lambda_R}
                                   +\theta_R  \Gamma_p\Gamma_n {\partial\over\partial\theta_R}
                                   \right] +\;
                             \nonumber{} \\  
 &
        +\;(\!(\lambda_L\theta_L\theta_L)\!)_{\alpha} \psi_L^{\alpha\mu}(x) {\partial\over\partial x^{\mu}}
        \;+\;
        \partial_{[n}\xi_{LRm]}^{\hat{\beta}}(x)[\theta_R^3\lambda_R]^{[mn]}
        \left(
              {\partial\over\partial\theta_R^{\hat{\beta}}}
              -
              (\Gamma^n\theta_R)_{\hat{\beta}}
              {\partial\over\partial x^n}
              \right)
        \;+
        \nonumber{} \\  
 &+\;\xiLLL^{\alpha [mn]}(x)(\!(\lambda_L\theta_L\theta_L)\!)_{\alpha}
                  \left(
                        \lambda_L\Gamma_{mn}{\partial\over\partial\lambda_L}
                        + \theta_L\Gamma_{mn}{\partial\over\partial\theta_L}
                        \right)     \;+\nonumber{} \\  
 &+\;\xiLRR^{\alpha mn}(x)(\!(\lambda_L\theta_L\theta_L)\!)_{\alpha}
                  \left( 
                        \lambda_R\Gamma_m\Gamma_n{\partial\over\partial\lambda_R}
                        + \theta_R\Gamma_m\Gamma_n{\partial\over\partial\theta_R}
                        \right)    \;+\ldots\nonumber{} \end{align}       
and a similar formula for $Q_R'$.

\subsection{Comparison to SUGRA}\label{sec:ComparisonToSUGRA}

The only fermionic superfields of \cite{Berkovits:2001ue} are $C^{\alpha\hat{\gamma}}_{\beta}$
and $C^{\hat{\alpha}\gamma}_{\hat{\beta}}$. The top component of $C^{\alpha\hat{\gamma}}_{\beta}$
corresponds to $\partial_{[m}\xi^{\hat{\gamma}}_{LRn]}(\Gamma^{mn})^{\alpha}_{\beta}$,
and the top component of $\hat{C}^{\hat{\alpha}\gamma}_{\hat{\beta}}$ to
$\partial_{[m}\xi^{\gamma}_{RLn]}(\Gamma^{mn})^{\hat{\alpha}}_{\hat{\beta}}$.

\section{Supersymmetries and dilatation}\label{SUSY}

The vector field $Q^{\rm flat}$ of Eq. (\ref{QFlat}) is manifestly supersymmetry-invariant.
In other words, it commutes with the super-Poincare algebra, which is generated by
${\partial\over\partial \theta_L^{\alpha}} - \Gamma^m_{\alpha\beta}\theta_L^{\beta} {\partial\over\partial x^m}$ and
${\partial\over\partial \theta_R^{\hat{\alpha}}} - \Gamma^m_{\hat{\alpha}\hat{\beta}}\theta_R^{\hat{\beta}} {\partial\over\partial x^m}$. It is also invariant under dilatations, if we define the weight of $x$
to be twice the weight of $\theta_L$, $\theta_R$. It is perhaps less straightforward to see that there
are no other symmetries. For example, there are no conformal symmetries. (But the dilatation symmetry is present.)
We will now prove that there are no other symmetries.

We have to compute the cohomology of $Q^{\rm flat}$ in the space of vector fields of ghost number $0$.
The cohomology of
$\lambda_L^{\alpha}{\partial\over\partial\theta_L^{\alpha}} + \lambda_R^{\hat{\alpha}}{\partial\over\partial\theta_L^{\hat{\alpha}}}$
at the ghost number $0$ is (see Section \ref{CohomologyQ0}):
\begin{align} T_m \;=\;
 &\partial\over\partial x^m\nonumber{} \\ S^L_{\alpha} \;=\;
 &\partial\over\partial \theta_L^{\alpha}\nonumber{} \\ S^R_{\hat{\alpha}} \;=\;
 &\partial\over\partial \theta_R^{\hat{\alpha}}\nonumber{} \\ D^L\;=\; 
 &\lambda_L^{\alpha}{\partial\over\partial\lambda_L^{\alpha}}
                   + \theta_L^{\alpha}{\partial\over\partial\theta_L^{\alpha}}\nonumber{} \\ M^L_{mn} \;=\;
 &\left(\lambda_L\Gamma_{mn} {\partial\over\partial \lambda_L}\right) +
                \left(\theta_L\Gamma_{mn} {\partial\over\partial \theta_L}\right)\nonumber{} \\ D^R\;=\; 
 &\lambda_R^{\hat{\alpha}}{\partial\over\partial\lambda_R^{\hat{\alpha}}}
                   + \theta_R^{\hat{\alpha}}{\partial\over\partial\theta_R^{\hat{\alpha}}}\nonumber{} \\ M^R_{mn} \;=\;
 &\left(\lambda_R\Gamma_{mn} {\partial\over\partial \lambda_R}\right) +
                \left(\theta_R\Gamma_{mn} {\partial\over\partial \theta_R}\right)\nonumber{} \end{align}
This means that any infinitesimal symmetry can be brought to the form:
\begin{align} v\;=\;
 &T^m(x){\partial\over\partial x^m} +\nonumber{} \\  
 &+\; D_L(x)\left(
                        \lambda_L^{\alpha}{\partial\over\partial\lambda_L^{\alpha}}
                        + \theta_L^{\alpha}{\partial\over\partial\theta_L^{\alpha}}
                        \right)
            +
            M_L^{mn}(x)\left(
                             \left(\lambda_L\Gamma_{mn} {\partial\over\partial \lambda_L}\right) +
                             \left(\theta_L\Gamma_{mn} {\partial\over\partial \theta_L}\right)
                             \right)
            +\nonumber{} \\  
 &+\; D_R(x)\left(\lambda_R^{\hat{\alpha}}{\partial\over\partial\lambda_R^{\hat{\alpha}}}
                                  + \theta_R^{\hat{\alpha}}{\partial\over\partial\theta_R^{\hat{\alpha}}}
                                  \right)
            +
            M_R^{mn}(x)\left(
                             \left(\lambda_R\Gamma_{mn} {\partial\over\partial \lambda_R}\right) +
                             \left(\theta_R\Gamma_{mn} {\partial\over\partial \theta_R}\right)
                             \right)
            +\nonumber{} \\  
 &+\; S_L^{\alpha}(x){\partial\over\partial \theta_L^{\alpha}}
            +
            S_R^{\hat{\alpha}}(x){\partial\over\partial \theta_R^{\hat{\alpha}}}\;+\ldots\nonumber{} \end{align}
where $\ldots$ stand for terms of the higher order in the grading defined by Eq. (\ref{GradingOperator}).
Commuting $v$ with $\left( (\!(\lambda_L\theta_L)\!)^m + (\!(\lambda_R\theta_R)\!)^m \right){\partial\over\partial x^m}$,
we have to cancel the coefficients of all generators of $[Q^{(0)}_L + Q^{(0)}_R\,,\,\_]$
(see Section \ref{sec:FirstPage}).
The vanishing of the coefficient of
$(\!(\lambda_R\theta_R)\!)^m \left(
                        \lambda_L^{\alpha}{\partial\over\partial\lambda_L^{\alpha}}
                        + \theta_L^{\alpha}{\partial\over\partial\theta_L^{\alpha}}
                        \right)$
implies that $D_L(x) = D_{L0}$ (constant in $x$). Similarly, $M^{mn}_L (x) = M^{mn}_{L0}$,
$D_R(x) = D_{R0}$, $M^{mn}_R (x) = M^{mn}_{R0}$.
The vanishing of the coefficient of
$(\!(\lambda_L\theta_L)\!)^m {\partial\over\partial x^n}$
and
$(\!(\lambda_R\theta_R)\!)^m {\partial\over\partial x^n}$ imply:
\begin{align}  
 &D_{L0} = D_{R0}\;=:D_0\nonumber{} \\  
 &M_{L0}^{mn} = M_{R0}^{mn}\;=:M_0^{mn}\nonumber{} \\  
 &T^m(x) = T^m_0 + 2D_{0}x^m + M_{0}^{mn} x^n\nonumber{} \end{align}
The vanishing of the coefficients of
$(\!(\lambda_R\theta_R)\!){\partial \over\partial\theta_L}$
and
$(\!(\lambda_L\theta_L)\!){\partial \over\partial\theta_R}$
imply $S_L^{\alpha}(x) = S_{L0}^{\alpha}$
and $S_R^{\hat{\alpha}}(x) =S_{R0}^{\hat{\alpha}}$ (do not depend on $x$).

\section{Acknowledgments}\label{Acknowledments}

This work was supported in part by  ICTP-SAIFR FAPESP grant 2016/01343-7,
and in part by FAPESP grant 2019/21281-4.
We want to thank Nathan Berkovits and Andrey Losev for useful discussions.

\appendix{}

\section{Higher spin conformal Killing tensors}\label{FiniteDimensionalSolutions}

Consider tensor fields on the flat $N$-dimensional space ${\bf R}^N$ with coordinates:
\begin{equation}
   x^m \quad,\quad m\in \{1,\ldots,N\}
   \end{equation}
They are functions with indices: $f_{m_1,\ldots m_r}(x)$, where $r$ is the rank of the tensor.
There are some differential equations which only have finite-dimensional spaces of solutions.
For example:
\begin{equation}
     {\partial\over\partial x^{(m}} f_{n)}(x) = 0
     \label{SymmetrizedDerivative}\end{equation}
The solutions of this equation are parameterized by constant antisymmetric tensors $b_{mn}$:
\begin{equation}
   f_m = b_{mn} x^n
   \end{equation}
More generally, consider the equation:
\begin{equation}
   {\partial\over\partial x^{(m_0}}f_{m_1\ldots m_n)}(x) = 0
   \label{GeneralEquation}\end{equation}
We want to classify the solutions of this equation.
Consider the Taylor expansion of $f_{m_1\ldots m_n}$ near $x=0$. Since Eq. (\ref{GeneralEquation}) is
homogeneous in $x$, we can consider each order of the Taylor expansion separately. In other words,
it is enough to consider $f_{m_1\ldots m_n}(x)$ a homogeneous polynomial of $x$. 
Let us introduce auxiliary variable $y^m$ and consider the generating function:
\begin{equation}
   \hat{f}(x,y) = y^{m_1}\cdots y^{m_n}f_{m_1\ldots m_n}(x)
   \label{HatFGeneratingFunction}\end{equation}
Homogeneous polynomials $\hat{f}(x,y)$ of $x,y$ of the order $N$ form a finite-dimensional representation
of $sl(2,{\bf R})$, with the generators defined as follows:
\begin{equation}
   E = y^m{\partial\over\partial x^m}\,,\;
   F = x^m{\partial\over\partial y^m}\,,\;
   H = y^m{\partial\over\partial y^m} - x^m{\partial\over\partial x^m}
   \end{equation}
Eq. (\ref{GeneralEquation}) implies that $\hat{f}(x,y)$ is a highest weigh vector:
\begin{equation}
   E\hat{f} = 0
   \end{equation}
On the other hand, $\hat{f}$ being a polynomial of the order $n$ in $y$ implies:
\begin{equation}
   F^{n+1}\hat{f} = 0
   \end{equation}
Therefore, the space of polynomial solutions of Eq. (\ref{GeneralEquation}) decomposes into the direct
sum of representations of dimension $0, 1, 2,\ldots, n$.
They correspond to polynomials of degree $0, 1, 2,\ldots, n$ in $x$.
We conclude that all solutions of Eq. (\ref{GeneralEquation}) are polynomials of order $n$ in $x$
(not necessarily homogeneous).

Let us now consider a weaker equation. Instead of requiring $\partial_{(m_0}f_{m_1\ldots m_n)}$ be zero,
we require the existence of $g_{m_2,\ldots m_n}(x)$ such that:
\begin{align}  
 &{\partial\over\partial x^{(m_0}}f_{m_1\ldots m_n)}(x) = \delta_{(m_0m_1}g_{m_2\ldots m_n)}(x)\label{GeneralEquationModuloDelta} \\  
 &\delta^{m_1m_2}f_{m_1\ldots m_n} \;=\; 0\label{Traceless} \end{align}
(We can think of Eq. (\ref{GeneralEquationModuloDelta}) as having a gauge symmetry
    $\delta f_{m_1\ldots m_n} = \delta_{(m_1m_2}h_{m_3\ldots m_n)}$,
    $\delta g_{m_2\ldots m_n} = \partial_{(m_2}h_{m_3\ldots m_n)}$,
    and Eq. (\ref{Traceless}) as fixing the gauge.)
The solutions of Eq. (\ref{GeneralEquationModuloDelta}) are higher spin conformal Killing tensors.
They correspond to traceless Killing tensors in AdS \cite{Mikhailov:2002bp}. Given a traceless
Killing tensor in AdS, we can consider the leading Taylor coefficient of its expansion around
a point in AdS. It will satisfy Eq. (\ref{GeneralEquation})
(with an additional condition $\delta^{m_1m_2}f_{m_1m_2\ldots m_n} = 0$)
implying that the space of solutions is finite-dimensional.

\def\cprime{$'$} \def\cprime{$'$}
\providecommand{\href}[2]{#2}\begingroup\raggedright\endgroup


\end{document}